\documentclass{emulateapj}
\usepackage{natbib,apjfonts}
\shorttitle{The core/cusp problem}
\shortauthors{de Blok}

\begin{document}

\title{The Core-Cusp Problem}

\author{W.J.G. de Blok}
\affil{Department of Astronomy, University of Cape Town, Rondebosch 7700, South Africa}

\begin{abstract}
  This paper gives an overview of the attempts to 
  determine the distribution of dark matter in low surface brightness
  disk and gas-rich dwarf galaxies, both through observations and
  computer simulations.  Observations seem to indicate an
  approximately constant dark matter density in the inner parts of
  galaxies, while cosmological computer simulations indicate a steep
  power-law-like behaviour. This difference has become known as the
  ``core/cusp problem'', and remains one of the unsolved problems in
  small-scale cosmology.
\end{abstract}

\section{Introduction}

Dark matter is one of the main ingredients of the universe. Early
optical measurements of the rotation of spiral galaxies indicated the
possible presence of large amounts of dark matter in their outer parts
(e.g.\ \citealt{Rubin:1978p737}), though in many cases the rotation
curve could also be explained by the stars alone (e.g.\
\citealt{Kalnajs:1983p87,Kent:1986p1363}). Observations at even larger
distances from the galaxy centers, using the 21-cm line of neutral
hydrogen, definitively confirmed the mass discrepancy
\citep{Bosma:1978p712,Bosma:1981p704,Bosma:1981p703}.  For an
extensive review see \citet{Sofue:2001p714}.  Most of these early
observations concentrated on late-type disk galaxies, which all share
the property of having an almost constant rotation velocity in their
outer parts (the so-called ``flat rotation curve''). As the dynamical
contribution of the stars and the gas is insufficient to explain the
high rotation velocities in the outer parts, this implies that most of
the observed rotation there must be due to some other material, the
``dark matter''. The observed constant velocity suggests that the dark
matter in the outer parts of galaxies has a mass density profile
closely resembling that of an isothermal sphere, i.e., $\rho \sim
r^{-2}$.

In the inner parts of galaxies stars are obviously present and they
must be the cause of a (possibly large) fraction of the observed
rotation velocity. This therefore leads to a transition from the inner
parts where the stars contribute to (and in many cases dominate) the
dynamics, to the outer parts where the dark matter is important (e.g.,
\citealt{Kalnajs:1983p87,Kent:1986p1363,Kent:1987p1410,vanAlbada:1986p701}).

The rotation velocity associated with dark matter in the inner parts
of disk galaxies is found to rise approximately linearly with
radius. This solid-body behaviour can be interpreted as indicating the
presence of a central core in the dark matter distribution, spanning a
significant fraction of the optical disk. Some authors adopt a
non-singular isothermal sphere to describe this kind of dark matter
mass distribution (e.g., \citealt{Athanassoula:1987p1446}), while
others prefer a pseudo-isothermal sphere (e.g.,
\citealt{Begeman:1991p1454,Broeils:1992p1452}).  Both models describe
the data well (see \citealt{Kormendy:2004p1455}), and by the late
1980s they had become the \emph{de facto} description of the
distribution of dark matter in (gas-rich, late-type) dwarfs and disk
galaxies.

In this paper we use the \emph{pseudo-isothermal} (PI) sphere to
represent the cored models, though this particular choice does not
affect any of the discussion in this paper.  The mass density
distribution of the PI sphere is given by:
\begin{equation}
\rho_{\rm PI}(r) = {{\rho_0}\over{1+(r/R_C)^2}},
\end{equation}
where $\rho_o$ is the central density, and $R_C$ is the core radius of the halo.
This density distribution leads to an asymptotic flat velocity $V_{\infty}$ given by
$V_{\infty} = (4\pi G \rho_0 R_C^2)^{1/2}$,
where $G$ is the gravitational constant.

In the early 1990s, the first results of numerical $N$-body
simulations of dark matter halos based on the collisionless cold dark
matter (CDM) prescription became available. These did not show the
observed core-like behaviour in their inner parts, but were better
described by a steep power-law mass-density distribution, the
so-called \emph{cusp}. Fits to the mass-distributions as derived from
these early simulations \citep{Dubinski:1991p4,Navarro:1996p479,
  Navarro:1997p482} indicated an inner distribution $\rho \sim
r^{\alpha}$ with $\alpha=-1$. (In the following we will use $\alpha$
to indicate the inner mass density power-law slope.)

The results from these and later simulations are based on the
$(\Lambda)$CDM paradigm, where most of the mass-energy of our universe
consists of collisionless CDM in combination with a cosmological
constant $\Lambda$.  This $\Lambda$CDM paradigm provides a
comprehensive description of the universe at large scales (as shown
most recently by the Wilkinson Microwave Anisotropy Probe (WMAP)
results; see \citealt{Spergel:2007p773}).   
However, despite these great successes, it should be kept in mind that the cusp
and the central dark matter distribution are not predicted from first
principles by $\Lambda$CDM. Rather, these properties are derived from
analytical fits made to dark-matter-only numerical simulations. While
the quality and quantity of these simulations has improved by orders
of magnitude over the years, there is as yet no ``cosmological
theory'' that explains and correctly predicts the distribution of dark
matter in galaxies from first principles.

The value $\alpha = -1$ found in the early CDM simulations is very
different from that expected in the PI model, where the constant
density core ($\rho \sim r^0$) implies $\alpha = 0$.  These two cases
thus lead to two very different descriptions of the dark matter
distribution in galaxies. The ``cusp'' ($\alpha=-1$) models gives rise
to a rapidly increasing ``spiky'' dark matter density towards the
center, while the ``core'' model ($\alpha = 0$) has an approximately
constant dark matter density. The cusp model therefore has a rotation
curve that will rise as the square-root of the radius, while the core
model rotation curve rises in a linear fashion. The difference in
shapes between the rotation curves of both models is quite pronounced,
and, in principle, it should therefore be possible to identify CDM
haloes in real galaxies by measuring their rotation curves.

Over the last 15 years or so, much effort has been put into
determining the central mass distribution in galaxies using their
rotation curves, and comparing them with the outcomes of ever more
sophisticated numerical simulations. To first order, one can summarize
this work as observational determinations yielding slopes $\alpha \sim
0$, while simulations produce $\alpha \sim -1$ slopes. This persistent
difference is known as the ``core/cusp controversy'', sometimes also
described as ``the small-scale crisis in cosmology''. The attempts to
reconcile the observations and simulations, either by trying to
improve them, or by trying to quantify systematic effects or missing
physics, are the subjects of this paper.  I give a brief overview of
past and present work dealing with the determination of the central
dark matter density distribution in galaxies, with an emphasis on the
observational efforts. An overview like this, touching on many
different topics in galaxy evolution, cosmology and computational
astrophysics, is never complete, and only a small (but hopefully
somewhat representative) fraction of the many papers relevant to this
topic can be referred to in the limited space available. The rest of
this paper is organised as follows: Sect.~2 gives a description of the
results that numerical simulations have produced over the
years. Section 3 deals with the observational determinations of the
dark matter density distribution. Section 4 discusses physical
scenarios that have been proposed to reconcile the core and cusp
distributions. Section 5 briefly summarizes the work discussed.

\section{Cold Dark Matter cusps}

The presence of a cusp in the centers of CDM halos is one of the
earliest and strongest results derived from cosmological N-body
simulations. \citet{Dubinski:1991p4} were among the first to
investigate the density profiles of CDM halos and found that the inner
parts of these simulated halos could be characterized by a power-law slope
$\alpha = -1$.  They did not rule out the existence of central cores,
but noted that these would have to be smaller than the resolution of
their simulations ($\sim 1.4$ kpc). Subsequent simulations,
at higher and higher resolutions, made the presence of cores in
simulated CDM haloes increasingly unlikely.

A systematic study by \citet{Navarro:1996p479, Navarro:1997p482} of
simulated CDM halos, derived assuming many different sets of
cosmological parameters, found that the innermost dark matter mass
density distribution could be well described by a characteristic
$\alpha = -1$ inner slope for all simulated halos, independent of
mass, size or cosmology. A similar general result was found for
the outer mass profile, with a steeper slope of $\alpha=-3$.
\citet{Navarro:1997p482} called this the ``universal density profile''
and it is described by
\begin{equation}
\rho_{\rm NFW}(r) = {{\rho_i}\over{(r/R_s)(1+r/R_s)^2}},
\end{equation}
where $\rho_i$ is related to the density of the universe at the time
of the time of halo collapse and $R_s$ is the characteristic radius of
the halo. This kind of profile is also known as the ``NFW
profile''.

The corresponding rotation curve is given by 
\begin{equation}
V(r) = \sqrt{ {{\ln (1+cx) - cx/(1+cx)}\over{x[\ln (1+c)-c/(1+c)]}}},
\end{equation}
with $x=r/R_{200}$. This curve curve is parameterized by a radius
$R_{200}$ and a concentration parameter $c=R_{200}/R_s$. Here $R_{200}$ is the
radius at which the density contrast with respect to the critical
density of the universe exceeds 200, roughly the virial radius;
$V_{200}$ is the circular velocity at $R_{200}$
\citep{Navarro:1996p479}. The parameters $c$ and $V_{200}$ are tightly
related through the assumed cosmology. Indeed, one can be expressed as
a function of the other, with only a small scatter
\citep{Bullock:2001p541}. That is, the range of $(c,V_{200})$
combinations that describes ``real'' CDM rotation curves is tightly
constrained by the $\Lambda$CDM cosmology.

Simulations by \citet{Moore:1999p517} indicated an even steeper inner
slope. They found that their simulated halos could be best described
by a function
\begin{equation}
\rho_{\rm M99}(r) = {{\rho_i}\over{(r/R_s)^{1.5}(1+r/R_s)^{1.5}}},
\end{equation}
i.e., with an inner slope $\alpha=-1.5$ and an outer slope
$\alpha=-3$.  

The difference between these two results indicated that issues such as
numerical convergence, initial conditions, analysis or interpretation
could still play a role in defining the inner slope.  As ever more
powerful computers and increasingly higher resolution simulations
became available, the value and behavior of the inner slope of CDM
halos has therefore been extensively discussed in the literature.
For example, to give but an incomplete listing of the many papers
that have appeared on this topic, \citet{Klypin:2001p46} derived
slopes $\alpha = -1.5$ for their simulated halos.  From phase-space
density arguments, \citet{Taylor:2001p684} argue that the density
profile should resemble an NFW profile, but converging to an inner
slope $\alpha = -0.75$, instead of the $\alpha = -1$ value.
\citet{Colin:2004p38} investigated low-mass haloes and found that they
were best described using NFW profiles (i.e., $\alpha=-1$).
\citet{Diemand:2005p13} found that CDM halos have cusps with a slope
$\alpha \simeq -1.2$.

Many studies assumed that the central cusp consisted of a region where
the mass density behaved as a power-law with a constant
slope. \citet{Navarro:2004p19} and \citet{Hayashi:2004p2} suggested
that this did not have to be the case. They did not find evidence for
an asymptotic power-law slope, but instead noted that the slope kept
getting shallower towards smaller radii without converging to a single
asymptotic value. At the smallest resolved radii they derive slopes of
$\sim -1.2$ for ``galaxy-sized'' halos (as measured at $\sim 1.3$ kpc), and $\sim
-1.35$ for ``dwarf galaxy'' halos (as measured at $\sim 0.4$ kpc). These values
are significantly steeper than the original NFW slope, but not as
steep as the \citet{Moore:1999p517} value.  \citet{Navarro:2004p19}
introduce a new fitting formula to quantify their results. For
reasonable choices of its input parameters, this formula yields an
extrapolated slope of $\alpha \sim -0.7$ at $r \sim 0.01$ kpc.

\citet{Stoehr:2006p25} also finds a gradual turn-over in slope towards
smaller radii.  Though his simulations formally resolve only radii
$\sim 1$ kpc (where a slope of $\alpha \sim -1$ is measured), an
extrapolation of his favoured fitting function towards smaller radii
results in a decreasing slope ending up as a flat slope ($\alpha=0$)
around $r \sim 0.01$ kpc.

\citet{Merritt:2005p605} and \citet{Graham:2006p616} showed that the
density distribution presented in \citet{Navarro:2004p19} could be
equally well described by a S\'ersic function. In the context of CDM
halos they refer to this function as an Einasto model. For
completeness, this profile is given by
\begin{equation}
\rho_{\rm Ein}(r) = \rho_e \exp\left( -d_n \left[ (r/r_e)^{1/n} -1 \right] \right),
\end{equation}
where $n$ determines the shape of the profile, and $d_n$ is a function
of $n$ which enables the use of the density $\rho_e$ measured at the
effective radius $r_e$. The latter is defined as the radius of
the volume containing half of the total mass. 
In terms of observationally more accessible quantities this can be written
as 
\begin{equation}
\rho_{\rm Ein}(r) = \rho_{\rm -2,Ein} \exp \left( -2n \left[ (r/r_{\rm -2,Ein})^{1/n}-1 \right] \right),
\end{equation}
where $r_{\rm -2,Ein}$ is the radius at which the logarithmic
derivative of the density profile equals $-2$, and $\rho_{\rm -2,Ein}$
is the density at that radius. The two versions of radius and density are related
by $\rho_{\rm -2,Ein} = \rho_e \exp(d_n-2n)$ and $r_{\rm -2,Ein} = (2n/d_n)^n r_e$.

A further discussion of this model is beyond the scope of this paper,
except to note that for typical parameterisations of CDM galaxy halos
one derives a slope of $\alpha \sim -1.3 \pm 0.2$ at a radius of 1 kpc
and of $\alpha \simeq -0.9 \pm 0.2$ at a radius of 0.1 kpc.  The
precise values depend on the exact values of $n$ and $r_{-2}$; the
values just listed assume $4<n<8$ and $r_{-2} = 10$ kpc, as shown in
\citet{Graham:2006p616}\footnote{Less steep slope values listed in the
  same paper are derived assuming $r_{-2} = 100$ kpc and are thus more
  appropriate for giant or group-sized CDM halos}.

Amongst the highest resolution measurements of the inner slope of CDM
halos so far are those by \citet{Navarro:2008p287} and
\citet{Stadel:2008p678}.  The former were done as part of the
Acquarius project \citep{Springel:2008p218}.  \citet{Navarro:2008p287}
found no convergence to a single asymptotic slope. Rather, as before,
the slope keeps decreasing with decreasing radius. At $r = 0.1$ kpc,
which is approximately the smallest reliably resolved radius, 
the slope has a value $\alpha \simeq 0.85$. At $r = 1$ kpc the value is
$\alpha \simeq 1.4$. An Einasto profile with $n=5.9$ provides a good
fit to the change in slope with radius.

In an independent, but equally detailed simulation,
\citet{Stadel:2008p678} find a similar behavior, as well as
comparable slope values. They quote a slope $\alpha = -0.8$ at 120 pc,
and $\alpha = -1.4$ at 2 kpc.

Even though the details of the simulations, the analytical fits and
the interpretation and analysis differ, we can still draw some
conclusions from the above discussion.  All simulations and fitting
functions considered here produce slopes $\alpha \la -1$ at a radius
of 1 kpc. At radii less than 1 kpc the most recent simulations tend to
produce slightly more shallow slopes where a typical value seems to be
$\alpha \simeq -0.8$ at 0.1 kpc.  From an observer's perspective, all
models described here produce slopes $\alpha \sim -0.8$ or steeper at
the smallest observationally accessible radii, and will thus all
produce results very similar to those derived using a ``standard'' NFW
profile. In the simulations, radii less than 0.1 kpc cannot yet be
reliably resolved, and the values of the slope derived there depend on
the validity of the assumed analytical fitting function.

\section{Observations}

\subsection{Early measurements}

The first comparisons of the HI rotation curves of gas-rich dwarf
galaxies with those
predicted by CDM profiles were presented in \citet{Moore:1994p86} and
\citet{Flores:1994p184}. The dynamics of these galaxies are dominated
by dark matter, and they are therefore thought to be good probes of
its distribution.  Both studies note a large discrepancy between the
observed rotation velocities and those predicted, especially in the
inner parts. They show that the PI model gives a superior description,
implying that the halos of these late-type dwarf galaxies are best characterized by an
approximately constant-density core. \citet{Moore:1994p86} briefly
addresses some of the observational uncertainties that might affect
the data, such as resolution, projection effects due to inclination,
and the effects of pressure support, and concludes they are not
significant enough to affect the results.  He also notes that it
is conceivable that, during the galaxy formation process, gas settling
in the halo will have affected the dark matter distribution. Usually
this is thought to take place in the form of a process called
``adiabatic contraction'' \citep{Blumenthal:1986p699}, which has the
effect of contracting the inner dark matter distribution (increasing
the density). If currently derived halo properties are the result of
this process, then the initial halos must have been of even lower
density, exacerbating the discrepancy.

\citet{Navarro:1996p192} argue that baryonic processes might be the
cause of the observed core distribution. They use $N$-body simulations
to model the effect of star formation on the baryons and the dark
matter, and find that a central dark matter core can be created if a
large fraction of the baryons is suddenly expelled into the halo. They
estimate that star formation rates of up to 10 $M_{\odot}$ yr$^{-1}$
are needed over a dynamical timescale of a galaxy for the process to
have the desired impact.

After analysing the same four dwarf galaxies that
\citet{Moore:1994p86} investigated, \citet{Burkert:1995p186} comes to
the conclusion that their rotation curves, after appropriate scaling,
are self-similar (see also \citealt{Salucci:2000p88}). He notes that
it is unlikely that baryonic blow-outs and mass-flows can cause this
kind of behaviour, unless fine-tuned, and attributes the slow rise of
the rotation curve to the intrinsic properties of the dark matter
(i.e., a dark matter core). He also finds tentative evidence that the
mass density in the outer parts of these dwarfs drops off as $r^{-3}$
(consistent with CDM), and not with $r^{-2}$ (as suggested by the
asymptotically flat rotation curves of spiral galaxies).  Based on
this, he introduces what has since become known as the
``Burkert-profile'':
\begin{equation}
\rho_{\rm Bur} = {{\rho_0 r_0^3}\over{(r+r_0)(r^2+r_0^2)}},
\end{equation}
with $\rho_0$ the central density, and $r_0$ the scale radius, similar
to the core radius $R_C$ of the PI model. This model is thus
characterised by $\alpha=0$ in the inner parts, and $\alpha = -3$ in
the outer parts.

A paper by \citet{Gelato:1999p191} presents a more detailed analysis
of the baryonic blow-out process and its impact on the halo
structure. They attempt to reproduce the observed rotation curve of
DDO 154 by simulating NFW halos, and subjecting them to the effect of
violent gas outflows, which are simulated by suddenly changing the
disk potential in the simulations.  In order for the final rotation
curve to resemble the observed curve, they need to suddenly blow out
33 to 75 percent of the initial disk material.  They note that they
can reproduce the rotation curve of DDO 154 for a rather wide range of
blow-out scenarios, and suggest that the fine-tuning argument
put forward by \citet{Burkert:1995p186} may not be applicable.

The blow-out process immediately gives rise to a number of
observational consequences. Firstly, a period of star formation
intense enough to blow out the majority of the baryons should leave
behind a substantial (by now) old stellar population. In practice,
however, these dwarfs are seen to be dominated by a young stellar
population. Is there a discrepancy here? Secondly, what happens to the
baryons that are blown out? Do they stay in the halo?  Presumably they
will be in the form of hot gas. Is this hot gas detectable?  Or, if
the gas cools down, can we see it raining back on the disk?
Discussing these questions in detail is beyond the scope of this
paper, but note that there are a number of starbursting dwarf galaxies
in our local universe where one can attempt to study these phenomena
directly.  As an example, \citet{Ott:2005p1541} analyse the properties
of the hot and cold gas in 8 dwarf galaxies that are in a starburst
phase. They show that outflows of hot gas are possible, but also that
the presence of (tidal) cold gas can again confine the hot gas. In the
galaxies in their sample, approximately 1 percent of the total ISM is
in the form of hot, coronal gas. The outflows are found to
be efficient in removing hot, metal-rich gas. Whether these 
processes can, in the early universe, also remove the bulk of the ISM
is still an open question.  It is clear that with this kind of
analysis we are no longer in the realm of cosmology, but are dealing
with ``messy'' astrophysics.

Fortunately, there is an alternative way to study the dark matter
distribution.  \citet{Navarro:1996p192} already note that the blow-out
process can only be effective in dwarf galaxies. In more massive
galaxies, such as spiral galaxies, the potential well is too deep to
efficiently remove the gas.  Finding and investigating more massive
dark-matter dominated galaxies may therefore be a more effective way
to explore the core/cusp issue. These galaxies, fortunately, do exist,
and are called Low Surface Brightness (LSB) galaxies.

\subsection{LSB Galaxies}

The term LSB galaxies is used here to indicate late-type, gas-rich,
dark-matter-dominated disk galaxies. Their optical component is
well-described by an exponential disk with an (extrapolated)
inclination-corrected central surface brightness fainter than
$\mu_{0,B} \sim 23$ mag arcsec$^{-2}$
\citep{McGaugh:1994p947,McGaugh:1995p958,deBlok:1995p952}.  Despite
their low surface brightness, their integrated luminosity is a few
magnitudes brighter than that of late-type dwarf galaxies ($M_B \sim -18$ to
$\sim -20$ for LSB galaxies, as opposed to $M_B \ga -16$ for the dwarf
galaxies). As noted, they are gas-rich ($M_{HI}/L_B \ga 1$; 
\citealt{Schombert:1992p1053,McGaugh:1997p1066,Schombert:2001p1060}), and
their interstellar medium has a low metallicity
\citep{McGaugh:1994p950,deBlok:1998p1076}. Their optical appearance is
dominated by an exponential disk with a young, blue population, with
little evidence for a dominant old population. Additionally, these
galaxies do not have large dominant bulges, and seem to have had a
star formation history with only sporadic star formation
\citep{vanderHulst:1993p956,vandenHoek:2000p1081,Gerritsen:1999p1104}. Central
light concentrations, if present at all, tend to be only fractionally
brighter than that of the extrapolated exponential disk.  In terms of
their spatial distribution, they are found on the outskirts of the
large scale structure filaments \citep{Bothun:1993p973,Mo:1994p949}.
In short, most observational evidence indicates that these galaxies
have had a quiescent evolution, with little evidence for major merging
episodes, interactions, or other processes that might have stirred up
the baryonic and dark matter (see also
\citealt{Bothun:1997p937,Impey:1997p1001}.

As for the term ``LSB galaxies'', there is some confusion in the
literature about what type of galaxies it applies to.  The type of LSB
galaxies most commonly studied, in particular with regards to the
core/cusp controversy, are the late-type LSB galaxies whose properties
are described above.  The other type of LSB galaxies often
discussed in the literature are the massive, early-type,
bulge-dominated LSB galaxies. These galaxies have properties entirely
different from the late-type LSB galaxies
\citep{Sprayberry:1995p977,Pickering:1997p988}.  The massive LSB
galaxies are a lot more luminous and their optical appearance is
dominated by a bright central bulge with a clearly detectable old
population \citep{Beijersbergen:1999p1156}. Many of them have
low-level AGN activity \citep{Schombert:1998p1225}. All indications
are that the evolution of these galaxies has been entirely different
from that of late-type LSB galaxies: if anything, they resemble S0
galaxies with extended disks, rather than late-type galaxies. The
presence of the dominant bulge also indicates that their central
dynamics are likely to be dominated by the stars, rather than dark
matter. In the following, the term ``LSB
galaxies'' therefore refers to late-type LSB galaxies only.

\subsection{Early HI observations of LSB galaxies}

The first detailed studies of large samples of LSB galaxies soon led
to the picture of them being unevolved, gas-rich disk galaxies, as
described above. The observation that they followed the same
Tully-Fisher relation as normal galaxies \citep{Zwaan:1995p951} was
intriguing, as this implied they had to be dark-matter
dominated. Follow-up radio synthesis observations in HI
\citep{deBlok:1996p5} soon confirmed this. Though the resolution of
these early observations was limited, the derived rotation curves
clearly resembled those of late-type dwarf and ``normal'' disk
galaxies: a slow rise, followed by a gradual flattening.  When
expressed in terms of scale lengths, the rotation curves of LSB and
HSB galaxies of equal luminosity turned out to be very similar,
indicating that LSB galaxies are in general low density objects
\citep{deBlok:1996p44}.

Mass models derived using the rotation curves clearly showed that for
reasonable assumptions for the stellar mass-to-light ratio,
$\Upsilon_\star$, the dynamics of LSB galaxies had to be dominated by
dark matter \citep{deBlok:1997p22}. Assuming that the stars had to
dominate the dynamics in the inner parts (the so-called maximum disk
solution) led to unrealistically high $\Upsilon_{\star}$ values, and,
even when taken at face value, still showed a need for a moderate
amount of dark matter at small radii (see also
\citealt{McGaugh:1998p34}).

The distribution of the dark matter at first sight seemed similar
to that in gas-rich dwarf galaxies \citep{deBlok:1997p22}.    Because
of the limited resolution of the data, \citet{deBlok:1997p22} did not
attempt fits with the NFW model, but noted that the halos had to be
extended, diffuse and  low density.

A first attempt at comparing the HI data with CDM predictions was made
by \citet{McGaugh:1998p34}. Rather than making fits to the rotation
curve, they simply assumed that the typical velocity $V_{200}$ of the
halo had to equal the outer (maximum) rotation velocity of the
galaxy. The strict cosmological relation between $c$ and $V_{200}$
then automatically yields a value of $c$ compatible with
$\Lambda$CDM. Adopting these values, the resulting halo rotation curve
turned out to be very different from the observed curve, in a similar
way as the \citet{Moore:1994p86} analysis: the NFW curve is too steep
and rises too quickly in the inner parts. The only way the halo curve
could be made to resemble the observed curve, was by abandoning the
cosmological $(c,V_{200})$ relation.

Similar conclusions were derived by \citet{Cote:2000p77}. They
presented high-resolution HI observations of dwarfs in the nearby
Centaurus and Sculptor groups, and noted that the derived rotation
curves did not agree with the NFW model.

A possible explanation that was soon put forward was that there were
still unrecognized systematic effects in the data, that would give the false impression of a 
core-like behaviour. Initially, attention focussed on the resolution
of the \citet{deBlok:1996p5} HI observations. These had beam sizes of
$\sim 15''$, resulting in the HI disks of the LSB
galaxies investigated having a diameter of 
between 3 and 18 independent
beams.  This limited resolution can potentially affect the shapes of the
rotation curves through a process called ``beam smearing'', as also 
mentioned in \citet{deBlok:1996p5}.  In observations with limited
resolution, the beam smearing process decreases the observed
velocities (compared to the true velocities), and in extreme cases can
turn any steeply rising rotation curve into a slowly rising solid-body
one.  This would therefore give the impression of a core being present
in the data, while the true distribution could still be cuspy. In their paper,
\citet{deBlok:1997p22} argued, through modelling of these beam
smearing effects, as well as the direct detections of steeply rising
rotation curves in the data, that while some beam smearing was indeed
present, the effect was not strong enough to completely ``hide'' the
dynamical signature of a cusp, and concluded that the data were
consistent with the existence of dark matter cores.

An alternative interpretation was given in
\citet{vandenBosch:2000p188} who used the \citet{deBlok:1997p22} data,
along with high-resolution literature rotation curves of a number of
late-type ``normal'' and dwarf galaxies, such as DDO154 and NGC
247. They derived and applied explicit analytical corrections for beam
smearing and concluded that the LSB galaxy HI data were consistent
with both cored and cuspy halos. In their analysis of the other
gas-rich dwarf and late-type galaxies, they found evidence for cores
in the dwarf galaxies, but detected a steep mass-density slope ---
consistent with a cusp --- in NGC 247, the one late-type disk galaxy
that met their sample selection criteria.

In a related paper, \citet{vandenBosch:2001p26} derive similar
conclusions for a different, larger sample of dwarf galaxies observed
in HI as part of the Westerbork HI Survey of Irregular and Spiral
Galaxies (WHISP) \citep{vanderHulst:2001p1232}.  They attempt to
correct for adiabatic contraction and resolution and conclude that,
while in a majority of the galaxies they investigate a core is
somewhat preferred in terms of fit quality, they cannot exclude halo
models that have a steep inner mass-density slope.

Clearly, with this wide range of (sometimes contradictory) conclusions,
obtaining higher resolution data is the only way to put stronger
constraints on the exact distribution of dark matter in galaxies.

\subsection{H$\alpha$ observations}

After the initial HI observations, the most efficient way to improve
the resolution was by obtaining observations in the H$\alpha$
line. These usually took the form of long-slit spectra taken along
the major axes of the galaxies, and resulted in an order of magnitude
improvement in resolution. Typical observations now had resolutions of
a few arcseconds, instead of a few tens of arcseconds.

Some of the first H$\alpha$ long-slit rotation curves of LSB galaxies
were presented by \citet{Swaters:2000p41}. They found that the
H$\alpha$ curves indeed did rise somewhat more steeply in the inner
parts than the HI curves, and for one or two galaxies very much so,
but they noted that on the whole the HI and H$\alpha$ curves were
consistent when the beam smearing effects were taken into account. The
H$\alpha$ curves and corresponding mass models did still indicate that
LSB galaxies had to be dominated by dark matter. The curves also still
rose less steeply than those of higher surface brightness late-type
galaxies of a similar luminosity, as indicated by the agreement
between the curves when they were radially scaled using their
exponential disk scale lengths.

A large set of high-resolution, H$\alpha$ long-slit rotation curves
was published and analysed by
\citet{McGaugh:2001p48}, \citet{deBlok:2001p65,deBlok:2001p84} and
\citet{deBlok:2002p47}.  Many of these galaxies had been part of the
\citet{deBlok:1997p22} sample, and a direct comparison between the
H$\alpha$ and HI rotation curves showed that in the majority of cases,
beam smearing effects were present, but not significant enough to alter
the previous conclusions regarding the dark matter distribution. In the few
cases where there were large differences this could be explained by
inclination effects or elliptical beam shapes in the HI observations.

The H$\alpha$ curves thus showed that the inner, slowly rising slopes
could not be caused by resolution effects. This meant that for
reasonable (stellar population synthesis inspired) $\Upsilon_\star$
values, LSB galaxies were still dark matter dominated throughout their
disk, even though the H$\alpha$ curves formally allowed maximum disk
fits with even higher $\Upsilon_{\star}$ values than the HI
observations did. These high values are, however, completely at odds
with the observed colours and star formation histories, and still
imply a significant dark matter fraction.

Comparison of mass models assuming PI and NFW models showed that the
data were better described by the PI models. In many cases, NFW models
did yield reasonable fits, but usually with very low concentrations
and high $V_{200}$ values, inconsistent with the 
cosmological $(c,V_{200})$ relation. Taken at face-value these results
would imply that the halos of LSB galaxies have barely collapsed, and
have typical velocities many times higher than those of the galaxies
that inhabit them.  The low $c$-values are, however,  due to the intrinsically
different shape of the NFW rotation curves compared to the solid-body
observed curves.  An NFW curve has an inner velocity slope $v
\propto r^{1/2}$, and the only way this kind of curve can fit the
observed solid-body $v \propto r$ curve is by stretching, 
resulting in the low $c$ and high $V_{200}$ values.

The observational inner mass density slopes were derived by \citet{deBlok:2001p65}
and plotted against the resolution (physical radius of the innermost point) of
the rotation curve. A comparison with the expected slopes from various
halo models showed that the majority of the data scattered towards the
predicted slopes of the PI model. They also showed that for
resolutions of $\sim 1$~kpc the PI and NFW models yielded identical
slopes. This would go some way towards explaining why some of the lower
resolution HI observations were unable to distinguish between models.
The behaviour of the change in slope when going from
the lower resolution HI observation to higher resolution H$\alpha$
observations is consistent with that expected from the PI
model, as shown in \citet{deBlok:2002p47}.
Clearly, for unambiguous determinations of the inner mass density
slope, resolutions of better than a kpc are needed.

Independent observations and analyses came to similar
conclusions. \citet{Marchesini:2002p78} obtained longslit H$\alpha$
observations of some of the \citet{deBlok:1997p22} galaxies, and also
found little evidence for strong beam smearing, as well as strong
evidence for the existance of dark matter
cores. \citet{Zackrisson:2006p1246} obtained long-slit observations of
the rotation curves of six extremely blue and bulge-less LSB galaxies
and also finds strong evidence for the presence of cores in the dark
matter distribution of these galaxies.  \citet{Salucci:2001p3}
analysed the rotation curves of over a hundred rotation curves of disk
galaxies, and found clear signatures for the existence of dark matter
cores in these galaxies. This analysis is complementary to the work on
LSB galaxies, as it also analyses less dark-matter-dominated galaxies.
 \citet{Borriello:2001p14} analysed
H$\alpha$ rotation curves of a number of dark matter dominated
galaxies, and also found strong evidence for core-like dark matter
distributions. They also fit Burkert haloes to their data and find
that these also provide good fits. Analysis for a sample of $\sim 25$
spiral galaxies is presented in \citet{Donato:2004p24}, and leads to
similar conclusions. They also find that the core radius of the dark
matter distribution is related to the disk scale length.

\citet{Hayashi:2004p2} find that the H$\alpha$ rotation curves
\emph{are} consistent with steep slopes.  \citet{deBlok:2005p449}
suggests that these conclusions are based on artificial constraints
imposed on the fitting functions. Once these constraints are removed,
17 of the 20 galaxies with the highest-quality data are best fitted
with cored models. Three are possibly consistent with cuspy models; two of
these are high-surface brightness dwarf galaxies that are likely 
dominated by stars.

\subsection{Possible systematic effects}

At first glance, the observational data seems to provide good evidence
for the presence of cores in LSB galaxies. This does of course not
exclude the possibility that systematic effects might be present in
the data that could give the (false) impression of cores. A number of
studies have therefore focused on these effects and asked whether the detected
cores could still be consistent with cuspy halo models.

The systematic effects that were investigated fall into two categories.

\begin{enumerate}
\item Pointing problems --- If small offsets exist between the central
  slit position and the true, dynamical center, this will cause the
  spectrograph slit to miss the cusp. This could be due to inaccurate
  telescope pointings or, alternatively, if large physical offsets
  between the dynamical and photometric centers of galaxies exist, the
  measured slopes will also be biased downwards.
\item
Non-circular motions --- The fundamental assumption in
the observational analyses discussed so far is that the gas moves on
circular orbits. If for some reason the orbits are elliptical, or of
the gas motions are disturbed, this will also lead to an
underestimate of the slope.
\end{enumerate}

These effects are difficult to recognize in long-slit, one-dimensional
H$\alpha$ data without additional information. Many authors therefore,
apart from modeling these effects, also emphasized the need for
high-resolution, two-dimensional velocity fields.  These make pointing
problems irrelevant, while non-circular motions can be directly
measured.  Fortunately, these velocity fields are now available (see
Sect.\ 3.6), largely superseding the results derived from the
H$\alpha$ rotation curves. Nevertheless, for completeness, the 
further analysis of the H$\alpha$ curves is briefly discussed here.

In \citet{deBlok:2003p11} a first attempt was made at modeling the
observational systematic effects.  Their main conclusion was that NFW
halos can be made to resemble dark matter cores only if, either
systematic non-circular motions with an amplitude of $\sim 20$ km
s$^{-1}$ exist in all disks, or systematic telescope pointing offsets
of $\sim 3-4''$  exist for all observations, or if the
dynamical and photometric centers are systematically offset in all
galaxies by $\sim 0.5-1$ kpc.

\citet{Marchesini:2002p78} and \citet{deBlok:2002p47} compare
independent sets of observations of the same galaxies, obtained at
different telescopes, by independent groups, using independent data
sets, and find no evidence for telescope pointing errors. In general,
galaxies can be acquired and positioned on the slit with a
repeatability accuracy of 0.3$''$ or so.

Using further modeling, \citet{deBlok:2003p11} find that a halo model
with a mildly cuspy slope $\alpha = -0.2 \pm 0.2$ gives the best
description of the data in the presence of realistic observational
effects. It is interesting to note that this best-fitting slope
had already been derived by \citet{Kravtsov:1998p159} on the basis of the
\citet{deBlok:1996p5} data.

An analysis by \citet{Spekkens:2005p36} of longslit H$\alpha$ rotation
curves of 165 low-mass galaxies comes to the same conclusion:
depending on how they select their sample, they find best-fitting
slopes of $\alpha = -0.22 \pm 0.08$ to $\alpha = -0.28 \pm 0.06$. They
also model pointing and slit offsets, but come to the conclusion that,
after correction, their data are consistent with cuspy haloes.
\citet{Swaters:2003p49} also present extensive modeling of
high-resolution long-slit H$\alpha$ rotation curves, and show that
while their data are consistent with $\alpha = 0$ cores, steeper
slopes cannot be ruled out.

Given the difference in interpretation of otherwise similar samples, a
double-blind analysis of the modeling performed by the various groups
would have been interesting.  However, with the
availability of high-resolution velocity fields, this is now a moot issue.

\citet{Rhee:2004p43} attempt to model many of the observational
effects using numerical simulations. Most of their conclusions apply
to long-slit observations. The systematic effects they investigate
(dealing with inclination effects, non-circular motions and profile
shapes) have now been directly tested on high-resolution velocity
fields and do not critically affect the data
(e.g., \citealt{Gentile:2005p89, Zackrisson:2006p1246,
  Trachternach:2008p152, Oh:2008p153, deBlok:2008p155,
  KuzioDeNaray:2008p50,KuzioDeNaray:2009p67}).

\subsection{High-resolution velocity fields}

As noted, high-resolution two-dimensional velocity fields provide the context which
long-slit observations are missing. With these velocity fields the
pointing problem becomes irrelevant, offsets between kinematical
and dynamical centers can be directly measured, as can 
non-circular motions.  Following is a brief overview of the various
observational studies that have presented and analysed these velocity
fields within the context of the core/cusp debate.

Some of the first high-resolution optical velocity fields of late-type
dwarf and LSB galaxies were presented by
\citet{BlaisOuellette:2001p1259}. They analyse H$\alpha$ Fabry-Perot
data of IC 2574 and NGC 3109, two nearby dwarf galaxies, and derive
slowly rising rotation curves, consistent with a core. This work was
later expanded in \citet{BlaisOuellette:2004p51}, and led to the work
presented in \citet{Spano:2008p143}, where optical velocity fields of
36 galaxies of different morphological types are presented.  All three
studies find that the PI model generally provided better fits than NFW
models. If NFW models give fits of comparable quality, then this is
usually at the cost of an unrealistically low $\Upsilon_{\star}$ value
and non-cosmological $(c,V_{200})$ values.  A similarly large
collection of H$\alpha$ velocity fields is presented in
\citet{Dicaire:2008p827}. They do not explicitly address the core/cusp
issue, but show that when bars are present, their influence on
the velocity fields is very noticable.

\citet{KuzioDeNaray:2008p50,KuzioDeNaray:2009p67} present DensePak
velocity fields of LSB galaxies, many of them taken from the
\citet{deBlok:1997p22} sample.  Their conclusions are that NFW models
provide a worse fit than PI models, for all values of
$\Upsilon_{\star}$. Where an NFW model could be fit, the $c$-values
generally again do not match the cosmological CDM $(c,V_{200})$ relation.
They introduce a ``cusp mass excess'': when the predicted
$(c,V_{200})$ relation is assumed, and the $V_{200}$ velocities are
matched with those in the outermost observed velocities, the inner
parts require about $\sim 2$ times more dark matter mass than is
implied by the observed rotation curves.

\citet{KuzioDeNaray:2008p50,KuzioDeNaray:2009p67} also explore
non-circular motions: they find that random velocities with an
amplitude of $\sim 20$ km s$^{-1}$ are needed to bring the observed
curves in agreement with the CDM predictions.  A comparison of
simulated long-slit observations (extracted from the velocity fields)
with the original long-slit data from \citet{McGaugh:2001p48} and
\citet{deBlok:2002p47} shows good
agreement.

\citet{Swaters:2003p79} present a DensePak velocity field of the late-type dwarf
galaxy DDO 39. They derive a rotation curve, and show that its slope
is steeper than implied by lower resolution HI data, and also
different from earlier long-slit data from
\citet{deBlok:2002p47}. They indicate that measurable non-circular
motions are present, but do not explicitly quantify them.  They show
fit results for NFW halos, but do not show the corresponding results
for a core model, so further comparisons are difficult to make.

\citet{Weldrake:2003p187} present the HI velocity field of NGC 6822, a
nearby Local Group dwarf galaxy. Their data has a linear resolution of
about $20$ pc, so beam smearing is definitely not a problem. The
rotation curve shows a strong preference for a core-like model, but
they do not quantify possible non-circular motions.
\citet{Salucci:2003p33} present similarly high-resolution HI data
(including VLA B array) of the dwarf galaxy DDO 47. Their analysis
shows the dynamics are not consistent with a cusp. Similar results are
derived in \citet{Gentile:2004p7} for a number of spiral galaxies.

\citet{Simon:2005p68} present results from CO and H$\alpha$ velocity
fields of 5 low-mass dark-matter-dominated galaxies (see also
\citealt{Simon:2003p52} and \citealt{Bolatto:2002p39}).  They derive a
range of slopes, from core-like ($\alpha = 0.01$ for NGC 2976) to
cuspy ($\alpha=1.20$ for NGC 5963). Note that NGC 5963 has an inner
bright disk, and it may therefore not be dark-matter-dominated all the
way to the center, making the value of its mass-density slope
uncertain (see also \citealt{Bosma:1988p1467}).  The average slope
they derive is $\alpha = 0.73 \pm 0.44$. Their analysis method differs
in a few aspects from the other studies referenced in this paper.
Firstly, their inner slope values are derived from a single power-law
fit to the entire rotation curve. Most of their models do not take into
account the gas component due to a lack of HI observations. In
low-mass galaxies the gas can dynamically be more important than the
stars (especially in the outer parts), and correcting for this
component could potentially change the derived slopes.

More importantly, the mass models in \citet{Simon:2005p68} are derived
under the explicit assumption of a constant inclination and position
angle for each galaxy. Most velocity fields of nearby galaxies show
radial inclination and position angle trends, especially in the outer
parts (e.g., \citealt{deBlok:2008p155}).  \citet{Simon:2005p68}
attribute these to radial velocities that give the
impression of changes in inclination and position angle.
Nevertheless, the non-circular velocities derived in this way are
typically less than $\sim 20$ km s$^{-1}$, and in a few galaxies less
than $\sim 5$ km s$^{-1}$.  Whether these velocities are real or
whether inclination and position angle changes are preferred would be
an interesting topic of further study.  \citet{Simon:2005p68} also
derive harmonic decompositions of the velocity fields to study
non-circular motions, but they do not list the values of the harmonic
coefficients. It would be similarly interesting to compare these with
results derived for other disk and LSB galaxies.

Direct measurements of  non-circular motions are presented in
\citet{Gentile:2005p89}. Using high-resolution HI observations, they
make a harmonic decomposition of the velocity field, and show that
non-circular motions are only present at the level of a few km
s$^{-1}$. This is a factor of $\sim 10$ lower than is needed for the
non-circular motions to wipe out the kinematical signature of a
cusp and to  give the impression of a core.

An analysis by \citet{Gentile:2007p90} of another gas-rich dwarf
galaxy, NGC 3741, based on the entire 3D HI data cube, showed
non-circular motions around 5$-$10 km s$^{-1}$, and a strong
preference for a core model. NFW models could be accommodated, but with the
usual caveat of the fit parameters not being consistent with the
cosmological $(c, V_{200})$ relation.

\citet{Trachternach:2008p152} present harmonic decompositions of the
velocity fields of galaxies from The HI Nearby Galaxy Survey
(THINGS; \citealt{Walter:2008p154}). The THINGS survey covers a large
range in galaxy properties, from luminous early-type disk galaxies to
late-type dwarfs.  \citet{Trachternach:2008p152} find a relation
between the median strength of the non-circular motions and the
luminosity of the galaxies, indicating that the non-circular motions
are associated with the baryons. Luminous disk galaxies have
non-circular motions up to $\sim 30$ km s$^{-1}$, mostly associated
with bars and spiral arms.  These then decrease rapidly to a level of
only a few km s$^{-1}$ for dwarf galaxies like DDO 154.  They also
found that offsets between photometric and kinematic centers were
typically $\sim 200$ pc or less.

This low level of non-circular motions seems inconsistent with
observations by \citet{Pizzella:2008p35}, who find significant
non-circular motions in a sample of 4 LSB galaxies. However, their
galaxies all contain bright bulges and are therefore probably more
representative of the class of giant LSB galaxies, rather than the
late-type ones discussed in this paper (in this regard see also
\citealt{Coccato:2008p62}).

Finally, \citet{McGaugh:2007p64} and \citet{Gentile:2007p83} show that
the core/cusp issue is not limited to the very inner parts of
galaxies. As mentioned before, the shape of the NFW curve is
fundamentally different from that of observed curves. One can try and
work out the implications in two ways: \citet{McGaugh:2007p64} match
the observed outer rotation velocities with those of corresponding
cosmological NFW halos (effectively identifying NFW halos that have a
similar dark matter density at these outer radii as real galaxies),
and find that these halos are too massive. Due to the cusp mass excess
their average density is then also a factor of 2--3 too high (see also
\citealt{deBlok:2008p155}). \citet{Gentile:2007p83} take a different
approach and identify halos that have the same enclosed mass as
observed galaxies. Again, due to the cusp mass excess, this results in
halos that, compared to what can be derived for real galaxies, are
more dense in the center and less dense in the outer parts. Due to the
fundamentally different shapes of the mass distributions, the
core/cusp issue is therefore not limited to the inner parts, but
is relevant at all radii.

\section{Effects of baryons and triaxiality}

The high-resolution velocity fields thus seem to indicate mass
distributions that are cored. As many observational effects such as
pointing, center offsets, and even non-circular motions seem to be too
small to be the cause of the observed cores, many studies have
attempted to explain the apparent presence of cores as the result of
processes such as interactions between and merging of dark matter
halos, or the effect of baryons on the dark matter
distribution. Following is a brief discussion of some of these
effects.

\subsection{Feedback and merging}

It was already noted that violent, large-scale star formation might be
able to explain the cores in gas-rich dwarf galaxies, but not in more
massive galaxies (Navarro et al 1996). The implied large bursts of
star formation are also not consistent with the quiescent evolution of
LSB galaxies. An alternative way to achieve the removal of the cusp
was proposed by \citet{Weinberg:2002p85}. They model a rotating rigid
bar in a disk which is embedded in a CDM halo and show that this bar
creates a wake in the dark matter.  This trailing wake slows down the
bar, transferring some of the bar angular momentum to the dark
matter. This ``puffs up'' the dark matter distribution thus forming a
core. Recent numerical simulation work by \citet{Dubinski:2009p1542}
does, however, suggest that the bar does not destroy the central cusp,
and may even increase the halo density slightly.

A very detailed case-study of the effects of bars and feedback is
presented in \citet{Valenzuela:2007p42}. They attempt to reproduce the
observed rotation curves of NGC 6822 and NGC 3109 using NFW halos and
a variety of feedback effects and non-circular motions.  They present
numerical models including gas dynamics, and tune them to resemble the
two target galaxies.  Whilst they succeed in matching the rotation
curves, it is not clear whether these results can be generalized, due
to, e.g., the different assumptions that are made in determining
inclinations and position angles of the models, compared to the data.
Although the authors present simulated velocity fields, they do not
make a direct comparison between observed and simulated velocity
fields, which would provide much additional information on the bar and
feedback effects.

\citet{Dekel:2003p74, Dekel:2003p6} argue that merging of cuspy halos
inevitably leads to a cusp. The only way to prevent this from
happening is to puff up the dark matter distribution of the infalling
halos before they merge. This way the halos get disrupted more easily
and a core-like distribution can be gradually built up. The authors
describe the scenario as speculative, however, and note that it is
unlikely that supernovae will be able to cause the puffing up for any
galaxy with $V>100$ km s$^{-1}$.  \citet{BoylanKolchin:2004p8} show
that for this process to work, none of the merging halos can be cuspy
to start with, as only mergers between cored halos give a cored merger
product. Any merger involving a cusp inevitably leads to a cuspy
end-product. \citet{Dehnen:2005p837} also shows that the final slope
always equals that of the steepest component. Setting up cored halos
and ensuring that they remain cored is apparently not trivial.

\subsection{Dynamical Friction}

\citet{ElZant:2001p56} propose a different way to make halos cored:
they note that merging gas clouds of $\sim 10^5\ M_{\odot}$ (for
dwarfs) to $\sim 10^8\ M_{\odot}$ (for spirals) can disrupt cusps
through dynamical friction. If this happens early enough in the
universe (when halos were smaller), this process could be very
efficient.  Similar scenarios are presented in \citet{Tonini:2006p59}.
\citet{RomanoDiaz:2008p70} in their study also argue that the effect
of the baryons on the halo structure must be significant.  They
suggest that in the presence of baryons, initially a very steep cusp
is formed (with $\alpha \sim -2$), which is then heated by sub-halos
through dynamical friction, and, subsequently, from the inside out
becomes shallower.  According to their analysis, the end result is a
density profile that is less steep than $\alpha = -1$ in the inner few
kpc, and may even be cored in the very center.

\citet{Jardel:2009p372} take the opposite view. They model the
dynamical friction process, but argue that it is difficult to find
``free-floating'' baryon clumps massive enough to make this process
happen, as these clumps must not be associated with dark matter (due to
the cusp that will then be formed; see above). They note that work by
\citet{Kaufmann:2006p1262} implies clump masses that are two orders of
magnitude too small.

The processes just described are similar to those proposed by
\citet{Mashchenko:2006p60, Mashchenko:2008p373}; see also
\citet{Ricotti:2004p28}.  They make it all happen in the early
universe, when mass scales and the required amount of baryons were both
smaller.  Their numerical simulations suggest that cusps can be erased
in the very early universe ($z\la 10$) when (proto-) galaxies had
approximately the size of the HI holes and shells observed in the disks of
present-day, gas-rich spiral and dwarf galaxies.  In objects that small, a
random motion of only $\sim 7$ km s$^{-1}$ (i.e., equal to the
HI velocity dispersion observed in local disk galaxies) is 
sufficient to disrupt the cusp and keep the halos cored until the
present day. Note though that in a recent analysis,
\citet{Ceverino:2009p374} perform similar calculations, but find that
halos remain cuspy, and suggest that differences in the simulation
approach might be responsible for this.

\citet{Chen:2008p375} also argue against the idea of creating cores at
high redshift.  Their argument is that cuspy halos cannot explain
gravitational lensing results, but demand halos with an even steeper
mass distribution (a singular isothermal sphere with slope $\alpha =
-2$). This means that if the halos of the elliptical galaxies that do
the lensing indeed form with NFW-like profiles, these need to steepen
over the course of their evolution, using some process for which the
quiescent adiabatic contraction is the most likely candidate.  LSB
galaxies, on the other hand, need to experience vigorous bursts of
early star formation that drive feedback to erase the initial
cusp.  These scenarios are at odds with what we know about the
evolution of these galaxies: ellipticals typically undergo a large
amount of merging, with the associated vigorous star formation, as
evidenced by their extensive old stellar populations.  LSB
galaxies show no evidence at all for any kind of violent interactions,
nor intense star formation. Dynamical friction and adiabatic
contraction thus seem to place demands on the evolution of elliptical and LSB
galaxies contrary to what can be derived from their star formation
histories.

\subsection{Triaxiality}

\citet{Hayashi:2006p97} and \citet{Hayashi:2007p94} show that
introducing an elliptical disturbance in an NFW potential can lead to
systematic (non-circular) motions that, when unrecognized, could be
interpreted as evidence for a core.  Their arguments particularly
apply to longslit H$\alpha$ observations, where indeed the context of
a velocity field is not available to gauge the validity of the
circular motion assumption. As described earlier, work on
high-resolution velocity fields by \citet{Gentile:2005p89},
\citet{Trachternach:2008p152} and \citet{KuzioDeNaray:2009p67} has put
significant observational constraints on the strength of non-circular
motions and the ellipticity of the (equatorial) potential.  For a
large sample of disk and dwarf galaxies, this potential is consistent
with being round and the non-circular motions are too small to give
the illusion of cores in long-slit observations.

The results from \citet{Hayashi:2006p97} and \citet{Hayashi:2007p94}
assume massless disks.  \citet{Bailin:2007p66} present results from
simulations of self-consistent massive disks in triaxial halos and
find that the baryons circularize the potential rapidly, even for
low-mass disks, thus wiping out a large part of the tri-axiality
signal.  \citet{Widrow:2008p58} also presents models of ``live'' disks
in tri-axial halos, and can approximate an observed LSB rotation curve
by introducing tri-axiality in the halo.

In all these studies it would be interesting to compare simulated
velocity fields with the observed ones. As has become clear from the
preceding discussion, modeling the long-slit rotation curves leaves
too many ambiguities which only studies of the velocity fields can
address.  In an observational study, \citet{KuzioDeNaray:2009p67}
subject model NFW velocity fields to the DensePak observing procedure,
and show that even in the presence of observational uncertainties the
signature of an NFW velocity field can be observed.

They show that axisymmetric NFW velocity fields are unable to
reproduce the observed velocity fields for any combination of
inclination and viewing angle. They derive NFW velocity fields in an
elliptical potential, and show that the only way these can be made to
resemble the observations is by having the observer's line of sight
along the minor axis of the potential for 6 out of the 7 galaxies
investigated, inconsistent with a random distribution of the line of sight.

\citet{KuzioDeNaray:2009p67} also note that the kind of rapidly varying
ellipticity of the potential as proposed by \citet{Hayashi:2007p94}
might help in better describing the data, but that the problem of a
preferred viewing angle will remain. For an elliptical potential with
a random viewing angle, one also expects, once in a while, to observe
a rotation curve that is steeper than the corresponding axisymmetric
NFW profile.  Rotation curves like that are, however,
exceedingly rare, if not absent, in the available observations.

\section{Summary}

Rotation curves of LSB and late-type, gas-rich dwarf galaxies indicate the presence of constant-density or mildy cuspy
($\alpha \sim -0.2$) dark matter cores, contradicting the predictions
of cosmological simulations. The most recent simulations still
indicate resolved mass density slopes that are too steep to be easily
reconciled with the observations (typically $\alpha \sim -0.8$ at
a radius $\sim 0.1$ kpc).  Claims of shallow slopes at even smaller radii
depend on the validity of the analytical description chosen for the mass-density profile.

Whereas early HI observations and long-slit H$\alpha$ rotation curves
still left some room for observational (pointing, resolution) or
physical (non-circular motions, tri-axiality) systematic effects to
create the illusion of cores in the presence of a cuspy mass distribution,
the high-resolution optical and HI velocity fields that have since
become available significantly reduce the potential impact of these
effects. Measured non-circular motions and potential ellipticities are
too small to create the illusion of a core in an intrinsically cuspy
halo.

This indicates either that halos did not have cusps to begin with, or
that an as yet not understood subtle interplay between dark matter and
baryons wipes out the cusp, where the quiescent evolution of LSB
galaxies severely limits the form this interplay can take. Adiabatic
contraction and dynamical friction yield contradictory results, while
models of massless disks in tri-axial halos result in preferred
viewing directions.  LSB galaxy disks, despite their low
$\Upsilon_{\star}$ values, are not entirely massless, and observations
and simulations will need to take this into account. Similarly, the
difficulties in reconciling a possible underlying triaxial potential
with the circularizing effects of the baryons also needs to be
investigated. In short, studies which, constrained and informed by the
high-quality observations now available, self-consistently
describe and model the interactions between the dark matter and the
baryons in a cosmological context are likely the way forward in
resolving the core/cusp problem.

\acknowledgements I thank the anonymous referees for the constructive comments. The work of WJGdB is based upon research
supported by the South African Research Chairs Initiative of the
Department of Science and Technology and National Research
Foundation.

\bibliography{mybib}

\end{document}